\documentclass[onecolumn,showpacs,floatfix,nofootinbib,11pt]{revtex4}
\usepackage{amssymb,amsmath}
\usepackage{subfigure}
\usepackage[latin1]{inputenc}
\usepackage[brazil]{babel}
\usepackage[dvips]{graphics,color}
\usepackage{epsfig}\usepackage{float}
\newcommand{\ba}{\begin{eqnarray}}
\newcommand{\ea}{\end{eqnarray}}

\newcommand{\bec}{\begin{center}}
\newcommand{\eec}{\end{center}}

\newcommand{\bea}{\begin{array}}
\newcommand{\ear}{\end{array}}

\newcommand{\bfr}{\begin{flushright}}

\newcommand{\efr}{\end{flushright}}

\newcommand{\bege}{\begin{equation}}
\newcommand{\enge}{\end{equation}}

\newcommand{\be}{\begin{equation}}
\newcommand{\ee}{\end{equation}}

\newcommand{\beq}{\begin{eqnarray}}\newcommand{\benu}{\begin{enumerate}}\newcommand{\enu}{\end{enumerate}}
\newcommand{\eeq}{\end{eqnarray}}

\newcommand{\bx}{\begin{pmatrix}}
\newcommand{\ex}{\end{pmatrix}}

\usepackage{amsfonts}

\begin{document}

\title{Two-branes with variable tension model and the effective Newtonian constant}

\author{J. M. Hoff da Silva}
\email{hoff@feg.unesp.br}\affiliation{UNESP - Campus de Guaratinguet\'a - DFQ, Av. Dr.
Ariberto Pereira da Cunha, 333 CEP 12516-410, Guaratinguet\'a-SP,
Brazil.}


\begin{abstract}
It is shown that, in the two brane time variation model framework, if the hidden brane tension varies according to the phenomenological E\"{o}tv\"{o}s law, the visible brane tension behavior is such that its time derivative is negative in the past and positive after a specific time of cosmological evolution. This behavior is interpreted in terms of an useful mechanical system analog and its relation with the variation of the Newtonian (effective) gravitational `constant' is explored.
\end{abstract}
\maketitle

\section{Introduction}

In the last years, several braneworld scenarios have acquired a lot of
attention of the scientific community. In fact, the possibility of solving
central problems in high energy physics -- as the hierarchy one -- seems to be quite attractive \cite{DV}. In particular, one of the models proposed by Randall and Sundrum (RS) \cite{PRI} is composed by two four-dimensional branes placed at the singular points of an $S^{1}/Z_{2}$ orbifold in an $AdS_{5}$ bulk. In this model, in order to solve the gravitational equations of the system, it is necessary a fine tuning between the brane tensions:
they must have the same strength, but with opposite signs. In this view, a negative brane tension in the compactification scheme is, indeed, necessary. In General Relativity, however, the presence of a negative tension object is inherently unstable.

It is well known that the projected gravitational constant depends linearly on the brane
tension \cite{JAP,MAR} times a positive factor, therefore it is necessary a positive tension
in order to guarantee the correct sign of the Newtonian constant. Usually, this problem is solved by allowing the distance between the branes to vary. In fact, the dynamics of the radion field engenders a new dependence between the brane tension and the Newtonian constant \cite{NOVO}.

It was demonstrated in Ref. \cite{VARTEN} that the possibility of time varying
brane tensions may lead to very peculiar scenarios in the two $3$-brane model. Actually, by using
the so-called braneworld sum rules \cite{SUM}, it is possible to derive in a consistent way the functional form of the brane tension variation function. The possibility of the time variation of the brane tension, in
the braneword context, was taken into account in previous works \cite{VAC,GER}. In Ref.
\cite{GER} a time variable brane tension entered the full gravitational projection equations
procedure. It was obtained a cosmological solution to a Friedmann brane highly compatible to the observable symmetries.

A crucial characteristic about a time variation of the brane tension is that it implies a time
variation on the gravitational Newtonian `constant', just by the reason outlined in the above paragraphs.
In fact, dealing with higher dimensional theories, one learns that the fundamental constants of the universe are those defined in the higher dimensions, leaving for the usual observed constants the possibility of variation \cite{JB1}. In this paper, we investigate another theoretical motivation to the time variation of $G$, without varying the size of the extra dimension. In practical terms, the time tension variation of a $3$-brane is connected with a time variation of the Newtonian constant, for a five dimensional bulk, by \be G \propto G_{5}\lambda(t),\label{gvar1}\ee where $G_{5}$ is the higher dimensional gravitational constant (assumed as a true constant) and $\lambda$ is the brane tension. The relation accomplished by Eq. (\ref{gvar1}) is classic and, as remarked, we are assuming that the size of the extra transverse dimension is already fixed (stabilized). In this way, we are able to contemplate the contribution of the time variation of $G$ coming {\it exclusively} from the brane tension.

The aim of this paper is to go one step further in the analysis of the brane tension variation developed in \cite{VARTEN}, obtaining an explicit function for $\lambda(t)$ via a second order differential equation resulting from the two-branes sum rules, when the hidden brane tension obeys an E\"{o}tv\"{o}s-like variation law \cite{EOT}. This approach appears as a new theoretical motivation for a variable Newtonian `constant', since the size of the extra dimension is assumed as stable, allowing, in this way, a variation on $G$ coming exclusively from the brane tension variation. The obtained differential equation allows (since it has second time derivatives) useful interpretation via a mechanical system analog (MSA). We contrast our solution with some experimental bounds on the variational rate of the Newtonian `constant' variation.

It may be useful a short list of physical inputs that lead to our analysis, as well as some obtained results: 1) Being the universe described as a brane, and taking into account its dramatic change of temperature during its cosmological evolution, we study a braneworld scenario with time variable brane tensions. By using the two-branes braneworld sum rules, it is possible to derive an useful and consistent differential equation obeyed by the brane tension. 2) When the hidden brane obeys the phenomenological E\"{o}tv\"{o}s variation law, the visible brane tension respects a very interesting variation law, having the following main characteristics: in early times it is governed by an exponential law with negative derivative. After enough time, its behavior is dominated by an E\"{o}tv\"{o}s law, just as the hidden brane. An analogy between this behavior and a simple mechanical system seems to be instructive in order to interpret the brane tension variation. 3) The time variation in the brane tension engenders, inevitably, a time variation in the effective gravitational `constant', even for frozen extra dimensions. Therefore, it is crucial to study the solution in view of the current experimental bounds. In fact, due to the specific form of the time variation, it is possible to achieve a rate of variation of $G$ negative in the past and positive after a specific time, say $t_{\ast}$. The value of $t_{\ast}$ may also be used to constraint the coupling of the hidden brane tension.

Let us summarize the main steps of the paper. In the next Section we reobtain the braneworld sum rules in a very general way. At the end of our derivation, we particularize to the $3$-brane embedded into a five dimensional bulk. This Section is somewhat a review Section, accomplishing the basic necessary mathematical tools. In Sec. III, we study the main aspects of our solution, calling special attention to the fractional rate of variation of the brane tension. It is also presented in this Section an analogy with a simple classical mechanical system, useful for better understanding of the variable tension profile, and some discussion on the available experimental data, concerning the observational bound of $G$. In the last Section we finalize by pointing out some criticisms and future branches of research.

\section{Preliminaries}

For the sake of completeness, in this Section we shall derive the consistency
conditions for braneworld models and apply them to the five-dimensional case. Then, after accomplishing the general framework, we implement the brane tension variation.

In a poor description, the implementation of the braneworld sum rules consists in an ingenious manipulation of the Einstein gravitational equation, taking into account two main aspects: the existence of branes (appearing in the energy-momentum tensor as singular sources) and the compactness and periodicity of the internal space (guaranteed by the $Z_{2}$ symmetry in the RS model, for instance). Usually the use of the braneworld sum rules is restricted to study the viability of a particular model. In fact, the goal of the sum rules is to provide, given a set of simple inputs (such as the number of branes and its dimensions, necessary in order to accommodate an specific model), a one parameter family of consistency conditions. Here, we shall use the braneworld sum rules to explore the time variation of the brane tensions. We will obtain a second order differential equation for its variation coming exclusively from sum rules. It is a two-sided situation: since the functional form of the variation will be obtained from the sum rules, it is mathematically consistent. Nevertheless, its derivation has nothing to say about the mechanism which leads to a variable tension. In more colloquial words, following the hint (given by the huge variation of temperature of the universe during its expansion) of a non fixed brane tension, we will only study the ``kinematics'' of the brane tension.

The early idea of the braneworld sum rules is presented in Ref. \cite{ELL}, nevertheless its modern fashion was first developed in Ref. \cite{KGL}. In this Section, we shall follow the approach given in Ref. \cite{LMW}, since it is quite general and comprehensive. We start with a $D$-dimensional bulk spacetime endowed with a non-factorizable geometry, characterized by a spacetime metric given by \ba
ds^{2}&=&\left.G_{AB}dX^{A}dX^{B}=W^{2}(r)g_{\alpha\beta}dx^{\alpha}dx^{\beta}+g_{ab}(r)dr^{a}dr^{b}\label{2},\right.
\ea where $W^{2}(r)$ is the warp factor, $X^{A}$ denotes the coordinates of the full $D$-dimensional spacetime, $x^{\alpha}$ stands for the $(p+1)$ non-compact coordinates of the spacetime, and $r^{a}$ labels the $(D-p-1)$ directions in the internal compact space. As will be noted, the internal space needs to satisfy a topological requirement: it must be compact without boundary. Otherwise, the sum rules loose some of their usefulness, but can be used as boundary conditions.

The first trick allowed by the spacetime described by (\ref{2}) is to dismember the $D$-dimensional Ricci tensor in terms of its lower dimensional partners by \cite{KGL} \begin{eqnarray}
R_{\mu\nu}&=&\bar{R}_{\mu\nu}-\frac{g_{\mu\nu}}{(p+1)W^{p-1}}\nabla^{2}W^{p+1},\label{dois}
\\
R_{ab}&=&\tilde{R}_{ab}-\frac{p+1}{W}\nabla_{a}\nabla_{b}W,\label{3}\end{eqnarray} where $\tilde{R}_{ab}$, $\nabla_{a}$ and $\nabla^{2}$ are respectively the Ricci tensor, the covariant derivative and the
Laplacian operator constructed by means of  the internal space metric $g_{ab}$ and $\bar{R}_{\mu\nu}$ is the Ricci tensor derived from $g_{\mu\nu}$. Denoting the three curvature scalars by
$R=G^{AB}R_{AB}$, $\bar{R}=g^{\mu\nu}\bar{R}_{\mu\nu}$ and
$\tilde{R}=g^{ab}\tilde{R}_{ab}$ we have, from Eqs. (\ref{dois})
and (\ref{3}), \ba
\frac{1}{p+1}\Big(W^{-2}\bar{R}-R^{\mu}_{\mu}\Big)=pW^{-2}\nabla
W\cdot\nabla W+W^{-1}\nabla^{2}W \label{4}\ea and \be
\frac{1}{p+1}\Big(\tilde{R}-R_{a}^{a}\Big)=W^{-1}\nabla^{2}W,\label{5}
\ee where $R^{\mu}_{\mu}\equiv W^{-2}g^{\mu\nu}R_{\mu\nu}$ and
$R^{a}_{a}\equiv g^{ab}R_{ab}$ (note that with such notation $R=R^{\mu}_{\mu}+R^{a}_{a}$).

The relation above are quite useful, since we can relate the right-hand side of (\ref{4}) and (\ref{5}) with a total derivative in the internal space coordinates. In fact, using \be
\frac{\nabla \cdot (W^{\xi}\nabla W)}{W^{\xi+1}}=\xi W^{-2}\nabla
W\cdot \nabla W+W^{-1}\nabla^{2}W \label{6},\ee where $\xi$ is an arbitrary parameter, one can express the left-hand side of (\ref{6}) by means of the partial traces of Eqs. (\ref{4}) and (\ref{5}) by
\ba
\nabla \cdot (W^{\xi}\nabla
W)&=&\left.\frac{W^{\xi+1}}{p(p+1)}[\xi\big(W^{-2}\bar{R}-R^{\mu}_{\mu}\big)
+(p-\xi)\big(\tilde{R}-R^{a}_{a}\big)].\right.\label{7}
\ea Now, admitting that the large scale bulk gravitational field is governed by General Relativity, i. e.,
\be R_{AB}=8\pi
G_{D}\Big(T_{AB}-\frac{1}{D-2}G_{AB}T\Big),\label{8} \ee where $G_{D}$ is the gravitational (truly) constant in $D$ dimensions, it is easy to verify that \ba
R^{\mu}_{\mu}=\frac{8\pi
G_{D}}{D-2}(T^{\mu}_{\mu}(D-p-3)-T^{m}_{m}(p+1)),\label{9}\\
R^{m}_{m}=\frac{8\pi
G_{D}}{D-2}(T^{m}_{m}(p-1)-T^{\mu}_{\mu}(D-p-1)),\label{10}\ea where $T^{\mu}_{\mu}=W^{-2}g_{\mu\nu}T^{\mu\nu}$, so that $T^{M}_{M}=T^{\mu}_{\mu}+T^{m}_{m}$. Inserting the above relations into the Eq. (\ref{7}) one arrives at
\ba \Bigg(\frac{D-2}{8\pi G_{D}}\Bigg) \nabla \cdot (W^{\xi}\nabla
W)&=&\left.
\frac{W^{\xi+1}}{p(p+1)}\Bigg(T^{\mu}_{\mu}[(p-2\xi)(D-p-1)+2\xi]\right.\nonumber\\&+&\left.
T^{m}_{m}p\,(2\xi-p+1)+\frac{D-2}{8\pi
G_{D}}[(p-\xi)\tilde{R}+\xi
\bar{R}W^{-2}]\Bigg).\right.\label{11} \ea

Now, if the internal space is compact --- or equivalently, if the fields are periodic in the internal space variables --- the integration of $\nabla \cdot (W^{\xi}\nabla W)$ over the internal space must vanish, i. e., \ba \oint \nabla \cdot (W^{\xi}\nabla W)=0.\label{van}\ea Therefore, the right-hand side of Eq. (\ref{11}) gives
\ba && \oint \left.
W^{\xi+1}\Bigg(T^{\mu}_{\mu}[(p-2\xi)(D-p-1)+2\xi]\right.\nonumber\\&+&\left.
T^{m}_{m}p\,(2\xi-p+1)+\frac{D-2}{8\pi
G_{D}}[(p-\xi)\tilde{R}+\xi
\bar{R}W^{-2}]\Bigg)= 0.\right.\label{12} \ea

Eq. (\ref{12}) provides an one parameter family of consistency conditions for braneworld models. Each particular choice of the $\xi$ parameter leads to a correspondent constraint. In practice, however, only a few choices bring interesting conditions. A specific model is implemented by setting the values of $D$, $p$, and the partial traces of the stress-tensor. In this paper we are particularly concerned with the five-dimensional bulk case, i. e., $D=5$ and $p=3$. With such specifications, the internal space is composed by only one extra (transverse) dimension, implying $\tilde{R}=0$. Apart from that, if the visible brane shapes our universe, it must reproduce the fact that $\bar{R}=0$ with an accuracy of $10^{-120}M_{Pl}^{4}$, being $M_{Pl}$ the four-dimensional Planck mass. Therefore, Eq. (\ref{12}) reduces simply to
\ba \oint W^{\xi+1}\Bigg(T^{\mu}_{\mu}+2(\xi-1)T_{m}^{m}\Bigg)=0.\label{13}\ea

Now, by attributing an energy-momentum tensor, it is possible to derive the physical conditions. The specification of a variable tension scenario enters in the partial traces, as all inherent characteristics of the branes.

\section{Variable tension}

In order to set a comprehensive stress-tensor, we considerer the following ansatz:

\ba T_{MN}=\frac{-\Lambda G_{MN}}{8\pi G_{5}}-\sum_{i}\Bigg\{T_{3}^{(i)}+\kappa_{1}^{(i)}\frac{\partial T_{3}^{(i)}}{\partial_{t}}+\kappa_{2}^{(i)}\frac{\partial^{2}T_{3}^{(i)}}{\partial_{t}^{2}}\Bigg\}P[G_{MN}]_{3}^{(i)}
\delta(r-r_{i})\label{14}.\ea

Let us make a few remarks on the above expression, first about its standard terms and then about its generalizations. In (\ref{14}), $\Lambda$ stands for the bulk cosmological constant, $T_{3}^{(i)}$ is the i$^{th}$ time-dependent brane tension, $P[G_{MN}]_{3}^{(i)}$ is the pullback of the metric to the i$^{th}$ $3$-brane (typically the spacetime metric evaluated on the brane world volume) and $\delta(r-r_{i})$ localizes the (singular) branes in the bulk. Note that the treatment of singular branes, in the light of Eq. (\ref{13}), is considerably simplified due to the presence of the $\delta$ function.

The constants $\kappa_{1}^{(i)}$ and $\kappa_{2}^{(i)}$ are, respectively, the coefficients of the first and second derivatives with respect to time of the i$^{th}$-brane tension. The 3-brane tension has unit of [(energy)/(length)$^{3}$], therefore the units of the couplings are $\kappa_{1}^{(i)}\thicksim$ (energy)$^{-1}$ and $\kappa_{2}^{(i)}\thicksim$ (energy)$^{-2}$. Their rate will be important for the consistency of the model and its values may be fixed by the consideration of a cosmological scenario with variable brane tension. We shall comment on that in the last Section. It is important to stress that the first and second derivative contribution to the stress-tensor are not an automatic consequence of a time dependent brane tension. Instead, we are assuming that such terms contribute to the stress-tensor in order to find a consistent tension variation through the use of the braneworld sum rules. There are, roughly speaking, two distinct approaches to deal with a variable tension brane. The brane tension may be understood as a field, whose kinetic and interaction terms contribute in a specific way to the stress-tensor. This fundamental and comprehensive picture is the one used in the scope of string theory (see \cite{CORD1}, for instance) and in the framework of supersymmetric branes \cite{CORD2}. On the other hand, the brane tension may also be treated as an intrinsic characteristic of the brane. In this case, since the tension is not a fundamental field, we are forced to plug a contribution in the stress-tensor, studying its consequences. This {\it down-up} approach is the one used mainly in braneworld scenarios (see \cite{GER}). We shall to keep our analysis in this last case.

It is easy to note that if the branes tensions are constants, Eq. (\ref{14}) reduces to the usual stress-tensor used to derive the braneworld sum rules \cite{LMW}. It is also important to remark that some aspects of the two brane variable tension model was derived in Ref. \cite{VARTEN} for $\kappa_{2}^{(i)}=0$. As will be clear, we will obtain the shape of the visible brane tension function by considering the hidden brane an E\"{o}tv\"{o}s brane, just as \cite{VARTEN}. Since for $\kappa_{2}=0$ the solution is an exponential function, the consideration of higher order derivatives in Eq. (\ref{14}) seems not to be so crucial. We shall, however, keep the second derivative term, in order to better interpret the physical scenario. In fact, in the next Section, it will be possible to understand several important aspects of the tension behavior by considering a MSA. In other words, the presence of the $\kappa_{2}^{(i)}$ constant (since it gives rise to the second derivative term) allows the determination of an useful Lagrangian of a MSA, complementing our understanding about the tension variation.

The partial traces of Eq. (\ref{14}) are given by
\ba T^{\mu}_{\mu}=\frac{-4\Lambda}{8\pi G_{5}}-4\sum_{i}\Bigg\{T_{3}^{(i)}+\kappa_{1}^{(i)}\frac{\partial T_{3}^{(i)}}{\partial_{t}}+\kappa_{2}^{(i)}\frac{\partial^{2}T_{3}^{(i)}}{\partial_{t}^{2}}\Bigg\}\delta(r-r_{i})
\label{15} \ea and \ba T_{m}^{m}=\frac{-\Lambda}{8\pi G_{5}}.\label{16}\ea Let us study the particular consistency condition arising from the $\xi=-1$ choice\footnote{It is well known that the $\xi=-1$ choice is particularly interesting in order to reproduce the so-called Randall-Sundrum-like model fine tuning \cite{KGL}.} in the relation (\ref{13}) and particularize our analysis to the two-branes case ($i=vis,hid$), where $vis$ denotes the visible brane and $hid$ the hidden one. Substituting Eqs. (\ref{15}) and (\ref{16}) in (\ref{13}) one obtains the following condition
\ba T_{3}^{vis}(t)+\kappa_{1}^{vis}\frac{dT_{3}^{vis}(t)}{dt}+\kappa_{2}^{vis}\frac{d^{2}T_{3}^{vis}(t)}{dt^{2}}
+(vis\rightarrow hid)=0,\label{17}\ea where the last term stands for the same first three terms for the hidden brane tension with its respectives $\kappa$'s. Note that for the standard (non variable) case, we reobtain the usual Randall-Sundrum-like fine tuning \be T_{3}^{vis}=-T_{3}^{hid}.\label{FT}\ee

Since there is not a fundamental approach which results in a specific form for $T_{3}^{vis}(t)$, we shall proceed as follows. By giving an input to $T_{3}^{hid}(t)$, from Eq. (\ref{17}), we obtain the compatible function for the visible brane tension. We shall keep the analysis for the phenomenological interesting case of an E\" otv\" os \cite{EOT,GER} hidden brane. In other words, we assume that \be T_{3}^{hid}(t)=-\lambda^{hid}t\label{18},\ee where $\lambda^{hid}$ is a positive constant with units of [energy/(length)$^{4}$]. Eq. (\ref{18}) plays an important role in the interpretation of the next subsection; by now, let us explain two points concerning to its assumption. First, the hidden brane tension is negative. It guarantees a positive tension to the visible brane which is necessary to reproduce the correct sign to the Newtonian constant (\ref{gvar1}) in the fixed branes case. In this vein one of the branes has a negative (and so, gravitationally problematic) tension, just as the RS model \cite{PRI}. The peculiarity here is that the brane with negative tension is the hidden one. Next, in order to justify the Eq. (\ref{18}), we note that the E\"otv\"os law was derived in terms of temperature \cite{GER}, not time. In its standard form it reads \cite{EOT} \be \chi=\zeta(T_{c}-T),\label{eot}\ee where $\chi$ is the fluid membrane tension, $\zeta$ a constant, $T$ the temperature and $T_{c}$ a critical temperature denoting the highest temperature for which the membrane exist. We are considering that, as least classically, it is possible to change the order parameter from temperature to time linearly. It is, in fact, a strong assumption, but the main problem, we believe, rests on the fact that there is not a known way to derive the time dependence of the hidden brane from first principles. Consequently, we need to look at some plausible physical input. Note that the physical interpretation encoded in both Eqs. (\ref{18}) and (\ref{eot}) is essentially the same: as the time goes by the universe cools, and the absolute value of the brane tension increases. If one is willing to accept this approach, it is easy to see that Eq. (\ref{17}) gives
\ba \frac{d^{2}T_{3}^{vis}(t)}{dt^{2}}+\frac{\kappa_{1}^{vis}}{\kappa_{2}^{vis}}\frac{dT_{3}^{vis}(t)}{dt}+
\frac{1}{\kappa_{2}^{vis}}T_{3}^{vis}(t)-\frac{\lambda^{hid}}{\kappa_{2}^{vis}}\bigg[t+\kappa_{1}^{hid}
\bigg]=0,\label{19} \ea whose solution is given by
\ba T_{3}^{vis}(t)&=&\left.\lambda^{hid}t+\alpha_{1}\exp{\Bigg(-\frac{t}{2\kappa_{2}^{vis}}\Big[\kappa_{1}^{vis}+\sqrt{
(\kappa_{1}^{vis})^{2}-4\kappa_{2}^{vis}}\Big]}\Bigg)\right.\nonumber\\&+&\left.\alpha_{2}\exp{\Bigg(\frac{t}{2\kappa_{2}^{vis}}\Big[-
\kappa_{1}^{vis}+\sqrt{(\kappa_{1}^{vis})^{2}-4\kappa_{2}^{vis}}\Big]}\Bigg),\right.\label{20} \ea apart from an absorbed constant\footnote{We remark, by passing, that such absorbed constant needs $k_{1}^{hid}>k_{1}^{vis}$ in order to be positive.}. So, in order to get a compatible solution it turns out that $(\kappa_{1}^{vis})^{2}\geq 4\kappa_{2}^{vis}$. It is necessary to remark that the visible brane tension is always positive for $\alpha_{1}>0$ and $\alpha_{2}>0$. The shape of $T_{3}^{vis}$ shows that it starts from a constant value $(\alpha_{1}+\alpha_{2})$ being, then, damped at early times by the exponential terms. After enough time, the first term of (\ref{20}) becomes dominant in the tension evolution. For large values of time, $T_{3}^{vis}(t)$ behaves according to an E\"otv\"os brane, just as the hidden brane but with opposite sign. Hence, the fine tuning is respected in this regime\footnote{Note, however, that this fact ($T_{3}^{vis}(t)=-T_{3}^{hid}(t)$ only for $t\gg$) is not problematic in this model, since it already came from a consistency condition.}. We shall discuss the general behavior respected by $T_{3}^{vis}(t)$ in the next subsection.

\subsection{A simple mechanical system analog}

It is insightful to think of $T_{3}^{vis}(t)$ as representing the position of a particle according to the variation of the parameter $t$ (time), in an analogy with a simple mechanical system governed by Eq. of motion (\ref{19}). It is possible to define a Lagrangian function, which reproduces all the conservative terms of the equation of motion, given by
\ba L\big(\dot{T}_{3}^{vis},T_{3}^{vis},t\big)=\frac{\big(\dot{T}_{3}^{vis}\big)^{2}}{2}-\frac{\big(T_{3}^{vis}\big)^{2}}
{2\kappa_{2}^{vis}}+\frac{T_{3}^{vis}\lambda^{hid}}{\kappa_{2}^{vis}}\big[t+\kappa_{1}^{hid}\big],\label{23} \ea where a dot means derivative with respect to the time. Thus, Eq. (\ref{19}) may be recast in the form
\ba \frac{d}{dt}\frac{\partial L}{\partial \dot{T}_{3}^{vis}}-\frac{\partial L}{\partial T_{3}^{vis}}=-\frac{\kappa_{1}^{vis}}{\kappa_{2}^{vis}}\dot{T}_{3}^{vis}.\label{24}\ea The associated momentum of $T_{3}^{vis}$ is given by \ba \pi = \frac{\partial L}{\partial \dot{T}_{3}^{vis}}=\dot{T}_{3}^{vis},\label{25}\ea and the `energy' function becomes
\ba E&=&\left.\dot{T}_{3}^{vis}\frac{\partial L}{\partial \dot{T}_{3}^{vis}}-L\right.\nonumber\\&=&\left.
\frac{\big(\dot{T}_{3}^{vis}\big)^{2}}{2}+\frac{\big(T_{3}^{vis}\big)^{2}}
{2\kappa_{2}^{vis}}-\frac{T_{3}^{vis}\lambda^{hid}}{\kappa_{2}^{vis}}\big[t+\kappa_{1}^{hid}\big].\right.\label{26}\ea

Now it is possible to look at the energy variation in time. From Eqs. (\ref{26}) and (\ref{24}) we have
\ba \frac{dE}{dt}=-\frac{\kappa_{1}^{vis}}{\kappa_{2}^{vis}}\big(\dot{T}_{3}^{vis}\big)^{2}-\frac{\lambda^{hid}}
{\kappa_{2}^{vis}}T_{3}^{vis}\label{27}.\ea The first term in the above equation may be understood as the energy loss due to the friction term, while the second term represents the damping in the energy due to the interaction between the branes (note that it is proportional to $\lambda^{hid}$). From the $T_{3}^{vis}(t)$ function, it is easy to see that there is a particular time, say $t_{\ast}$, such that $\dot{T}_{3}^{vis}(t_{\ast})=0$. Hence, at $t_{\ast}$ the energy loss is only due to the interaction term. Since $T_{3}^{vis}(t)>0$ and the friction term is quadratic, the energy is monotonically decreasing as the time evolves. The friction term is the dominant one in the early behavior of $T_{3}^{vis}$. In fact the accelerated damping until $t_{\ast}$ corroborate it. The loss of energy for asymptotic times $(t\gg t_{\ast})$ is dominated by the uniform interaction term. So, the `particle' does not have enough energy to spend it changing its position in an accelerated way. Note also that the `particle', subject to the potential \ba V=\frac{(T_{3}^{vis})^{2}}{2\kappa_{2}^{vis}}-\frac{\lambda^{hid}T_{3}^{vis}}{\kappa_{2}^{vis}}(t+\kappa^{hid})
,\label{pot}\ea reach its minimum at $T_{3}^{vis}=\lambda^{hid}(t+\kappa_{1}^{hid})$, which match (apart from the constant $\kappa_{1}^{hid}\lambda^{hid}$) the asymptotic behavior of the visible brane tension. So, $T_{3}^{vis}$ is naturally led to the minimum of its potential. This minimum is, in fact, stable since the second derivative of the potential is $1/\kappa_{2}^{vis}$. This analogy calls our attention, once more, to the necessity of a positive $\kappa_{2}^{vis}$ parameter.

In the next subsection we shall complement our analysis addressing some issues on experimental bounds on the time variation of the Newtonian constant and its relation to the brane tension variability.

\subsection{Observational constraints}

The recent astrophysical data suggests the constancy of the Newtonian gravitational constant. Actually, the data are responsible to give a quite small upper bound in a possible time variation of $G$. As we mentioned, time dependence on the brane tension engenders time dependence on $G$. So, the solution $T_{3}^{vis}$ must respect these upper bounds. The new characteristic of the model is that the existence of the second derivative term in $T_{3}^{vis}$ gives rise to an unusual change in the $\dot{G}/G$ sign.

In order to further explore the aforementioned relation we need, initially, to look at the first derivative of $T_{3}^{vis}$. This is because the experimental bounds on the variability of the Newtonian constant are taken with respect to its fractional variation, i. e., $\dot{G}/G$. From Eq. (\ref{20}) it is easy to see that
\ba \dot{T}_{3}^{vis}(t)=\lambda^{hid}-\alpha_{1}A_{1}e^{-A_{1}t}-\alpha_{2}A_{2}e^{-A_{2}t},\label{21}\ea where $A_{1}\equiv \frac{1}{2\kappa_{2}^{vis}}\Big[\kappa_{1}^{vis}+\sqrt{(\kappa_{1}^{vis})^{2}-4\kappa_{2}^{vis}}\Big]$ and $A_{2}\equiv \frac{1}{2\kappa_{2}^{vis}}\Big[\kappa_{1}^{vis}-\sqrt{(\kappa_{1}^{vis})^{2}-4\kappa_{2}^{vis}}\Big]$. From Eqs. (\ref{20}) and (\ref{21}), and given the positivity of the $\kappa$'s parameters, it is easy to observe that while $\dot{T}_{3}^{vis}$ may be positive or negative, $T_{3}^{vis}$ is always positive. In particular, for $t=t_{\ast}$ the rate $\big[\dot{T}_{3}^{vis}/T_{3}^{vis}\big]\mid_{t=t_{\ast}}$ vanishes. With the aid of Eq. (\ref{21}), $t_{\ast}$ constraint the brane coupling $\lambda^{hid}$, in terms of the parameters $\kappa$'s and the integration constants (which shall be fixed by initial conditions) $\alpha$'s, by \be \lambda^{hid}=\sum_{j=1}^{2}\frac{\alpha_{j}A_{j}}{e^{A_{j}t_{\ast}}}\label{28}.\ee
The general picture emerging from $\dot{T}_{3}^{vis}/T_{3}^{vis}\big(=\dot{G}/G\big)$ as the time evolves is the following. Before $t_{\ast}$, $\dot{T}_{3}^{vis}<0$ and then $\dot{T}_{3}^{vis}/T_{3}^{vis}<0$. However, the fractional variation change its sign. In fact, for $t>t_{\ast}$ we have $\dot{T}_{3}^{vis}/T_{3}^{vis}>0$. As expected, if $t\sim t_{\ast}$ it is observed a rate of variation consistent with zero. It is an important outcome of this toy model. The dynamics of the brane tension provides a non-uniform time variation, in such a way that $\dot{T}_{3}^{vis}$ may change its sign.

Over the past last years some important experiments were carried out in order to probe a possible change of $G$ with the cosmic evolution of the universe (for a review see Ref. \cite{CW} and references therein). Here, we would like to enumerate some present values for the current upper bound on the variation of $G$. We refer the reader to Ref. \cite{UZAN} for a more complete table of measurements. Strictly speaking, the variation of a dimensional constant has no physical meaning, since its variation may be incorporated in a suitable redefinition of length, time and energy \cite{novo}. It is possible, however, to pick up an arbitrary unity and define it as the standard one. There are several observational constraints coming from bounding a difference between two values on $\dot{G}/G$. Assuming $G\sim t^{-\alpha}$, the analysis of big-bang nucleosynthesis \cite{NUCLEO} gives $(0\pm 4)\times 10^{-13}$ $yr^{-1}$, while bounds from helioseismology sets $(0\pm 16)\times 10^{-13}$ $yr^{-1}$ \cite{HELIO}. The study of the binary pulsar PSR 1913+16 provides $(40\pm 50)\times 10^{-13}$ $yr^{-1}$ \cite{PULSAR}. It is important to stress that these bounds are obtained in the context of a specific $G$ variation or a particular model. The best model independent bound on the $\dot{G}/G$ value is given by lunar laser ranging measurements: $(4\pm 9)\times 10^{-13}$ $yr^{-1}$ \cite{LUNAR}. In the present model, since the function $\dot{T}_{3}^{vis}/T_{3}^{vis}$ has a maximum shortly after $t^{*}$ and rapidly approaches the $t$-axis, it is possible to restrict the parameters requiring that \be \dot{T}_{3}^{vis}(\bar{t})/T_{3}^{vis}(\bar{t})<10^{-13} yr^{-1},\label{II1}\ee being $\bar{t}$ the solution of the transcendental equation $\ddot{T}_{3}^{vis}(\bar{t})T_{3}^{vis}(\bar{t})=[\dot{T}_{3}^{vis}(\bar{t})]^{2}$. In this way, the observational constraints are not violated at any time interval. Obviously, while the restriction (\ref{II1}) is sufficient not to contradict the experimental limits, it is clearly insufficient to put a boundary on each parameter individually.

It is quite useful to study, as an example, how the restriction (\ref{II1}) works more explicitly in the particular case where $\alpha_{1}=0$. Denoting $\alpha_{1}\equiv \alpha$ and $A_{2}\equiv A$ it is easy to see that Eq. (\ref{28}) reads \be \lambda^{hid}=\alpha A e^{-At_{*}}\label{III1}.\ee Therefore, $T_{3}^{vis}$ and its first derivative may be recast in a suitable way as \be T_{3}^{vis}(t)=\alpha e^{-At_{*}}\Bigg[At+\frac{e^{At_{*}}}{\alpha A^{2}}\ddot{T}_{3}^{vis}(t)\Bigg]\label{III2}\ee and \be \dot{T}_{3}^{vis}(t)=\alpha A e^{-At_{*}}\Bigg[1-\frac{e^{At_{*}}}{\alpha A^{2}}\ddot{T}_{3}^{vis}(t)\Bigg],\label{III3}\ee being $\ddot{T}_{3}^{vis}(t)=\alpha A^{2}e^{-At}$. Now, from the equation defining $\bar{t}$ \Big($\ddot{T}_{3}^{vis}(\bar{t})T_{3}^{vis}(\bar{t})=[\dot{T}_{3}^{vis}(\bar{t})]^{2}$\Big) we arrive, after some simple algebra, at \be e^{A(t_{*}-\bar{t})}=\frac{1}{A\bar{t}+2}.\label{III4} \ee Using Eqs. (\ref{III4}) and (\ref{II1}) it follows that \be \frac{1}{\bar{t}+A}<10^{-13} yr^{-1}, \label{III9} \ee making explicit the constraint which must be respected.

The present experimental bounds on $\dot{G}/G$ do not excludes the peculiar $\dot{T}_{3}^{vis}/T_{3}^{vis}$ change of sign. On the other hand, in view of the same present data it is impossible, up to now, to guarantee that $G$ is in fact a dynamical (and with a measurable dynamics) quantity. A promising possibility to scrutinize $\dot{G}$, at intermediary redshifts, comes from the study of gravitational waves. Being $G$ a variable quantity, there exists a subtle modification in the gravitational waves frequencies and phase of a inspiraling binary \cite{GW} which could, in principle, be measured by Space-borne gravitational wave detectors, such as the Laser Interferometer Space Antenna (LISA) detector. In light of Eqs. (\ref{20}) and (\ref{21}), however, it seems that the change in the $\dot{G}/G$ sign should be explored via the study of the $\dot{G}$ sign over a wide range of redshifts.

\section{Final Remarks}

In view of the huge variation of the temperature of the universe during its evolution attested by the CMB data, there is no reason to insist on a constant brane tension \cite{GER,VARTEN}. We studied a toy two-branes model presenting time variable tensions. As it is well known, from the projection of geometric quantities from the bulk to the brane, the effective gravitational constant depends linearly on the brane tension \cite{JAP,MAR,GER}, for a frozen extra dimension. Therefore a variable tension leads, inevitably, to a dynamical Newtonian `constant' even in a scenario without change in the size of the extra dimension.

Using the braneworld consistency conditions applied to the case in question, we were able to derive a second derivative differential equation whose solution gives the $T_{3}^{vis}(t)$ profile. The use of the consistency conditions is two-handed: on the one hand, it provides a strong mathematical result, completely consistent with the physical inputs. It is, obviously, the entire point of the braneworld sum rules. On the other hand, it is almost inescapable the use of some ad hoc assumptions in order to implement a specific model. In the present case, it is encoded in Eq. (\ref{18}). Even being this case the phenomenologically interesting one \cite{EOT} and already studied in several aspects in the literature \cite{GER}, here it enters as an additional hypothesis, rendering the obtained model the status of a toy model. The important feature, however, is that this simple model presents a new theoretical motivation to the $G$ variation, with a particular observational signature (the change in the $\dot{G}/G$ sign as the time evolves), which is not excluded by the present data.

As remarked, we arrived at a $G$ variation even by considering the extra dimension stabilized, i.e., working with a fixed extra dimensional size. This type of consideration is appealing since it allows the study of $\dot{G}/G$ coming only from the brane tension. Up to our knowledge, however, there is not a mechanism which stabilizes the compactification radius in the context of a variable brane tension. Thus, this assumption needs to be further investigated. In fact, it is well known that a weak brane tension complicates the modulo stabilization mechanism \cite{GWISE}. So, a variable brane tension contemplating both weak and strong regimes is certainly a difficult additional problem in the stabilization of the distance between the branes.

It is important to make a few remarks on the hierarchy problem in the context of variable tension branes. In the standard RS setup, the brane tensions have opposite signs of those we are dealing here\footnote{Again, note that in the context of fixed branes, a positive brane tension (as we consider here) is necessary in order to reproduce the right sign of the Newtonian gravitational constant.}. Therefore, at first, the hierarchy problem is not solved in the visible brane. We stress, however, that the case studied here does not have (do not need, for our purposes) an explicit metric, hence the warp factor form is completely open. Therefore, the question about the usefulness of variable tension branes with respect to the solution of the hierarchy problem must be further investigated. Perhaps, an extension of the analysis performed in Ref. \cite{VAC} to the two brane case shall shed some light on this open issue.

We finalize by point out that a variable tension also leads to a new degree of freedom in the gravitational field equations. From the cosmological point of view, such type of scenarios may present interesting features. It was shown in Ref. \cite{LAS2}, for instance, that a cosmological brane whose tension presents an E\"{o}tv\"{o}s-like variation law may remove the initial singularity of the universe. This bounce-like behavior is also present in the two-branes model \cite{VARTEN}. Other physical implications to cosmological models based upon variable brane tension may be founded in Refs. \cite{OUT}. Further explorations show that a variable brane tension, varying with the exponential of the scale factor, may describe matter creation in a cosmological framework \cite{COSMOL}. In this vein, it would be interesting to analyze the cosmological implications of variable brane tension, when the brane tension is given by a mixing of exponential and E\"{o}tv\"{o}s-like terms, just as Eq. (\ref{20}) prescribes. Such models may be useful in order to constraint the parameters which appear in the visible brane tension solution.

\section*{ACKNOWLEDGMENTS}

I would like to thank to professor Roldao da Rocha for valuable suggestions. It is also a pleasure to thank to professor Marcelo B. Hott for stimulating discussion.

\end{document}